# Equivalent Circuit Modeling and Analysis of Metamaterial Based Wireless Power Transfer

[1]Webster Adepoju, *Student Member, IEEE*, [1]Indranil Bhattacharya, *Member, IEEE*
[2]Ismail Fidan, [1]Nasr Esfahani Ebrahim, [1]Olatunji Abiodun, [2]Ranger Buchanan,
[1]Trapa Banik [1]Muhammad Enagi Bima, *Student Member, IEEE*
[1]Department of Electrical and Computer Engineering, [2]Department of Manufacturing and Engineering Technology
Tennessee Technological University
Cookeville, 38505, TN, USA
woadepoju42@tntech.edu, ibhattacharya@tntech.edu

*Abstract*—In this study, an equivalent circuit model is presented to emulate the behavior of a metamaterial-based wireless power transfer system. For this purpose, the electromagnetic field simulation of the proposed system is conducted in ANSYS high-frequency structure simulator. In addition, a numerical analysis of the proposed structure is explored to evaluate its transfer characteristics. The power transfer efficiency of the proposed structure is represented by the transmission scattering parameter. While some methods, including interference theory and effective medium theory have been exploited to explain the physics mechanism of MM-based WPT systems, some of the reactive parameters and the basic physical interpretation have not been clearly expounded. In contrast to existing theoretical model, the proposed approach focuses on the effect of the system parameters and transfer coils on the system transfer characteristics and its effectiveness in analyzing complex circuit. Numerical solution of the system transfer characteristics, including the scattering parameter and power transfer efficiency is conducted in Matlab. The calculation results based on numerical estimation validates the full wave electromagnetic simulation results, effectively verifying the accuracy of the analytical model.

*Index Terms*—Wireless Power Transfer (WPT), Finite Element Analysis (FEA), Scattering parameter, Metamaterial, Power Transfer Efficiency (PTE), HFSS, MatLab.

## I. INTRODUCTION

OVER the past years, Wireless Power Transfer (WPT) has demonstrated tremendous capability as a power transmission mechanism in mobile computing [1]–[3], wireless charging of biomedical body implants [4], [5], consumer electronics and wireless charging of Electric Vehicles (EV) [6], [7]. However, despite its increasing penetration, reliability concerns stemming from high power dissipation, leakage Electromagnetic Field (EMF), and low power transmission efficiency (PTE) for wide range WPT systems remain largely unresolved. To this end, attention has shifted to metamaterial (MM) as a viable alternative for range enhancement and performance improvement in WPT systems. Negative-index MMs (NIMs) are the first and most investigated of all MM structures [8] and typify a generic description for left handed materials, having a negative refractive index, $n$, as depicted in (1) (electric permittivity, $\epsilon_f = -1$, and magnetic permeability, $\mu_f = -1$), and hence supports perfect lensing [9], [10].

$$n = -\sqrt{\mu_f \epsilon_f} \qquad (1)$$

When operated at resonance, electromagnetic ($EM$) fields are confined inside the resonators, leading to a periodic exchange of electric and magnetic energy. Outside the resonators, however, the EM fields decay evanescently and do not carry away energy unless coupled to the tail of the evanescent wave of another resonator [11]. With NIM, the amplitude of evanescent waves can be enhanced such that the distance between two resonators are virtually smaller. The strategy for NIM design borders on reconstructing and modulating its structural properties, including effective permeability ($\mu_f$), and effective permittivity ($\epsilon_f$) coupled with a tuning of design parameters such that the reconstructed MM exhibits left-handed characteristics ($\mu_f < 0$ and $\epsilon_f < 0$).

Concretely, the evanescent wave amplification property of the proposed MM is of significant interest in this study. This is explained by the fact that magnetic resonant coupling depends on evanescent wave amplification of near magnetic field. Bearing in mind that inductive WPT system and evanescent wave amplifier utilize only the magnetic field for power transfer, it therefore goes without saying that a negative real part of effective permeability ($\mu_f < 0$) is a satisfactory requirement for the design under test to achieve evanescent wave amplification and negative refractive index [12]. To this end, this paper presents an equivalent circuit model to explore the transfer characteristics of an MM-based WPT system. While several methods, such as interference theory [13], [14], transmission line circuit model [15], [16], and effective medium theory [17] have been widely exploited in literature to explain the physics mechanism of MM-based WPT system, some of the reactive parameters and the basic physical interpretation have not been clearly expounded. In contrast to existing theoretical model, the proposed methodology focuses on the effect of system parameters and transfer coils on the transfer characteristics of the system which is pertinent in analyzing complex systems. Specifically, a four-coil structure comprising the drive coil ($dr$), transmitting coil ($tx$), receiving coil ($rx$), and load coil ($l$) is utilized for the analysis. Further, a performance comparison of the four coil WPT structure with the insertion of MM and without MM is analyzed both analytically and in finite element simulation, the MM being posited between the transmitting and receiving coils. Specifically, the main

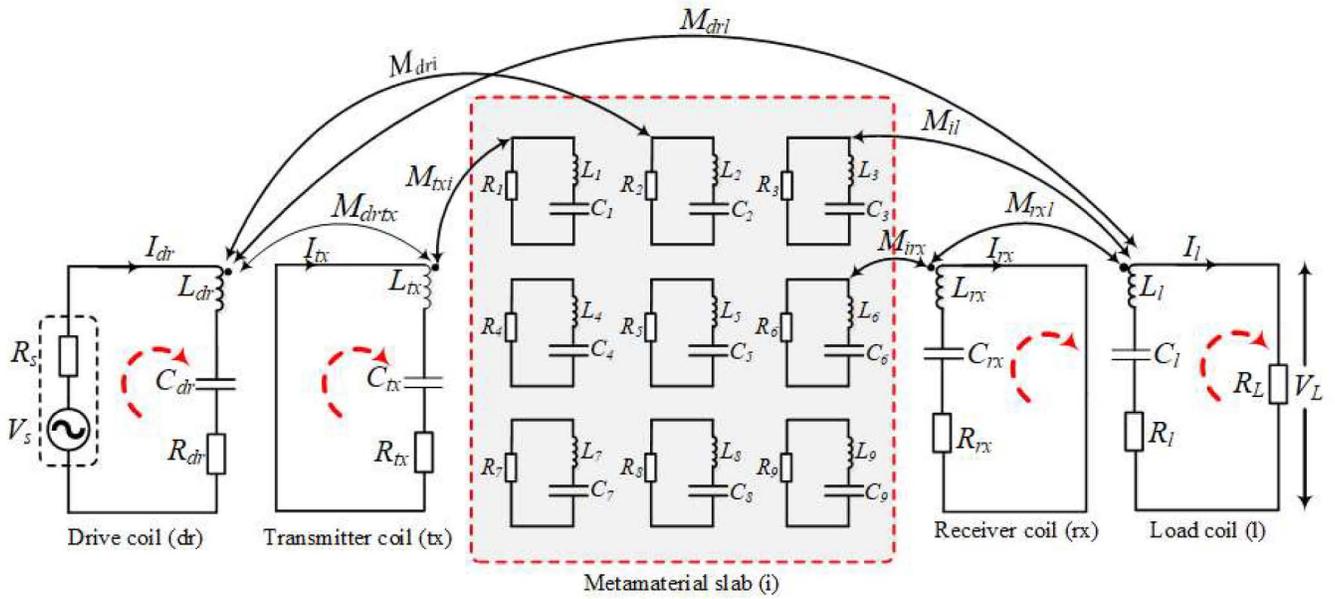

Fig. 1. Equivalent circuit schematic of the proposed four coil MM-based WPT System. The metamatarial slab comprises a 3×3 periodic array of MM unit cell in Fig. 2

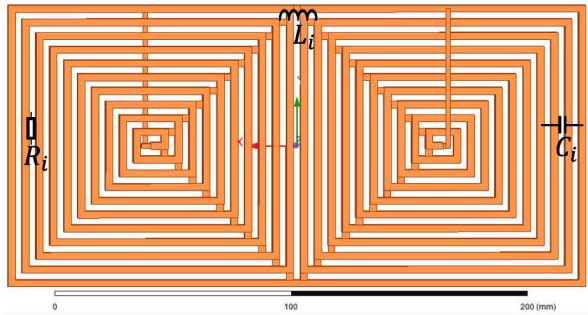

(a) Schematic of the proposed unit MM structure

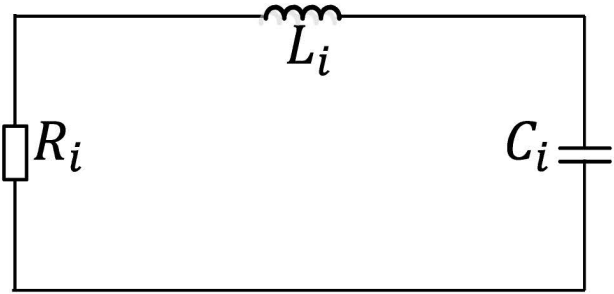

(b) Equivalent lumped resonant RLC circuit of the proposed MM in Fig. 2(a)

Fig. 2. Proposed MM unit cell and equivalent lumped circuit model

contributions of this manuscript are:
1) to derive an equivalent circuit model of a MM based wireless power transfer system
2) to determine the analytical solution of the transmission scattering parameter and power transfer efficiency of an MM-based WPT system
3) to present a performance comparison of the numerical solution and finite element simulation of the proposed equivalent circuit model.

The MM slab is composed of a 3×3 periodic array of the MM unit cell in Fig. 2(a). By leveraging printed circuit board technology, the unit cells are designed and fabricated into a planar structure. Afterwards, the planar cells are assembled into an evenly spaced 3×3 periodic array of the proposed unit cell, otherwise referred to as MM slab.

## II. SYSTEM DESCRIPTION, MODELING AND ANALYSIS

### A. Equivalent Circuit Modeling of the Proposed Metamaterial based WPT System

In order to simplify the equivalent circuit model of the proposed structure, the circuit schematic of proposed MM unit cell is modeled into coupling coils as seen in Fig. 2(a). Based on [18], [19], each unit cell can be modeled as a resistance-inductor-capacitor (RLC) resonant circuit exhibited in Fig. 2(b) where $L_i(i = 1, 2, 3....9)$, $C_i(i = 1, 2, 3....9)$, and $R_i(i = 1, 2, 3....9)$ are the lumped circuit inductance, capacitance and resistance of the unit cell, respectively. It is worth stating that the subscript $i$ represents each unit cell in the MM-slab. As shown in (2), the resistive component consists of the ohmic loss ($R_o$) and dielectric loss ($R_d$) due to the copper coil

$$R_i = R_o + R_d \qquad (2)$$

Similarly, the capacitive component is modeled as the sum of the stray capacitance ($C_s$) and compensation capacitor $C_{com}$ as exhibited in (3)

$$C_i = C_s + C_{com} \qquad (3)$$

It is worth stating that the resonant operating frequency of the MM-based WPT structure can be adjusted by varying the value of the compensation capacitor, $C_{com}$, accordingly. Based on the aforementioned, the performance of the proposed MM-based WPT system can be effectively optimized by tuning the value of $C_{com}$. Further, the MM-slab located between the transmitting and receiving coil in Fig. 1 is designed based on a 3×3 periodic array of the proposed unit cell in Fig. 2(a). Intuitively, such a loosely disconnected array of cells can be considered as a multiple repeater coils or booster network for achieving evanescent wave amplification of near magnetic field as well as increasing the amount of power transfer. Besides, the current flow through each unit cell of the MM-slab can be assumed identical and in phase while the overall system can be represented as a lump circuit model, in which the distribution of the parameters are ignored. Moreover, the parameters of the MM-based four coil WPT system in Fig. 1 is expounded below

⋄ $L_{dr}$, $C_{dr}$ and $R_{dr}$ are the self inductance, capacitance and resistance of the driver coil, respectively.
⋄ $L_{tx}$, $C_{tx}$ and $R_{tx}$ are the inductance, capacitance and resistance of the transmitting coil, respectively.
⋄ $L_{rx}$, $C_{rx}$ and $R_{rx}$ are the self inductance, capacitance and resistance of the receiving coil, respectively.
⋄ $L_l$, $C_l$ and $R_l$ are the self inductance, capacitance and resistance of the load coil, respectively.
⋄ $L_i$, $C_i$ and $R_i$ are the self inductance, capacitance and resistance of the load coil, respectively; where $i = 1, 2, 3 ..... 9$.
⋄ $M_{drtx}$ is the mutual inductance between the driver coil and transmitting coil.
⋄ $M_{dri}$ is the mutual inductance between the driver coil and each MM-slab.
⋄ $M_{txi}$ is the mutual inductance between the transmitting coil and MM-slab.
⋄ $M_{irx}$ is the mutual inductance between the MM-slab and receiving coil.
⋄ $M_{il}$ is the mutual inductance between the MM-slab and load coil.

Meanwhile, $Z_{dr}$, $Z_{tx}$, $Z_{rx}$, $Z_l$ and $Z_i$ denote the impedance of the driver coil, transmitter coil, receiver coil, load coil and each MM unit cell, respectively. Similarly, $I_{dr}$, $I_{tx}$, $I_{rx}$, $I_l$ and $I_i$ depicts the current flowing through the driver coil, transmitter coil, receiver coil, load coil and each MM unit cell, respectively. By leveraging the concept of coupling theory and Kirchhoff's voltage law (KVL), the transfer characteristics of the WPT system can be analyzed as demonstrated in (4).

$$\begin{cases} Z_{tx} = R_{tx} + j\left(\omega L_{tx} - \dfrac{1}{\omega C_{tx}}\right) \\ Z_{rx} = R_{rx} + j\left(\omega L_{rx} - \dfrac{1}{\omega C_{rx}}\right) \\ Z_{dr} = R_{dr} + j\left(\omega L_{dr} - \dfrac{1}{\omega C_{dr}}\right) \\ Z_l = R_L + R_l + j\left(\omega L_L - \dfrac{1}{\omega C_L}\right) \\ Z_i = R_i + j\left(\omega L_i - \dfrac{1}{\omega C_i}\right) \end{cases} \quad (4)$$

where $\omega = 2\pi f_{res}$ and $f_{res}$ is the resonant operating frequency. Further, the current flowing through each of the resonant circuit and MM unit cell can be evaluated by applying Kirchoff's Voltage Law (KVL) to the equivalent circuit in Fig. 1.

$$V_s = I_{dr}(R_s + Z_{dr}) + j\omega M_{drtx}I_{tx} + j\omega M_{dr1}I_1 + j\omega M_{dr2}I_2 \\ + \ldots j\omega M_{dr9}I_9 + j\omega M_{drtx}I_l \quad (5)$$

$$0 = I_{tx}Z_{tx} + j\omega M_{drtx}I_{dr} + j\omega M_{tx1}I_1 + j\omega M_{tx2}I_2 + \\ \ldots j\omega M_{tx9}I_9 + j\omega M_{txL}I_L \quad (6)$$

$$0 = I_iZ_i + j\omega M_{dri}I_{dr} + j\omega M_{tx1}I_1 + j\omega M_{tx2}I_2 + \\ \ldots j\omega M_{tx9}I_9 + j\omega M_{il}I_i \quad (7)$$

$$0 = I_l(R_L + Z_l) + j\omega M_{dri}I_{dr} + j\omega M_{txl}I_{tx1} + j\omega M_{tx2}I_2 + \\ \ldots j\omega M_{tx9}I_9 + j\omega M_{il}I_i \quad (8)$$

Converting the expressions (5), (6), (7) and (8) to matrix form gives

$$\begin{bmatrix} V_S \\ 0 \\ 0 \\ 0 \\ \vdots \\ 0 \end{bmatrix} = \begin{bmatrix} R_s + Z_{dr} & jwM_{drtx} & jwM_{dri} & \cdots & jwM_{drl} \\ jwM_{drtx} & Z_{tx} & jwM_{txl} & \cdots & jwM_{txl} \\ jwM_{dri} & jwM_{txi} & Z_1 & \cdots & jwM_{il} \\ \vdots & \vdots & \vdots & & \vdots \\ jwM_{drl} & jwM_{tl} & jwM_{iI} & \cdots & R_L + Z_l \end{bmatrix} \begin{bmatrix} I_{dr} \\ I_{tx} \\ I_i \\ \vdots \\ I_l \end{bmatrix} \quad (9)$$

Given the weak magnetic coupling between two non-adjacent coils, the associated mutual inductance can be effectively neglected as depicted in (10).

$$M_{dri} = M_{txrx} = M_{il} = M_{drl} = 0 \quad (10)$$

Notice that the essential mutual inductances are determined from the finite element simulation software as described in section III. The calculation results are obtained based on the theoretical analysis of (4), (5), (6), (7), (8) and (9). In addition, the essential components (capacitor, inductance and resistance) obtained in accordance with [20]. Using the four KVL equations in (4) and the circuit matrix in (9) and leveraging the numerical method presented in [6] in Matlab simulation environment, the resulting voltage transfer characteristics are obtained as in (11)

$$\frac{V_L}{V_S} = \frac{\omega^3 L_{tx}L_{rx}R_L\sqrt{L_{dr}L_l}}{Z_{dr}Z_{tx}Z_{rx}Z_l + \omega^2(L_{dr}L_{rx}Z_{rx}Z_l + L_{tx}L_{rx}Z_{dr}Z_l \\ + L_{rx}L_lZ_{dr}Z_{tx}) + \omega^4(L_{dr}L_{tx}L_{rx}L_l)} \quad (11)$$

In agreement with [6], [20] coupled with the expression in (11), the equivalent transmission scattering parameter, $S_{21}$, is evaluated as shown in (12)

$$S_{21} = 2\frac{V_L}{V_S}\left(\frac{R_S}{R_L}\right)^{\frac{1}{2}} \quad (12)$$

The corresponding power transmission efficiency (PTE) is then calculated as shown in (13)

$$PTE = |S_{21}|^2 \times 100\% \quad (13)$$

In order to achieve impedance matching and maximum PTE, it is assumed that the source resistance, $R_s$, and load resistance ($R_l$) are 50Ω [13]. In practice, however, the impedance matching can be realized by a simple adjustment of the distance between the drive coil ($dr$) and transmitter ($Tx$) as well as between the receiver ($Rx$) and load coil ($l$).

### III. FINITE ELEMENT ANALYSIS OF THE PROPOSED MM-BASED WPT SYSTEM

In order to ascertain the accuracy of the proposed equivalent circuit, the power transfer efficiency and system forward transmission coefficient are investigated based on a comparative analysis of the theoretical model and finite element simulation. Fig. 3 depicts a 3-dimensional design of the proposed equivalent circuit.

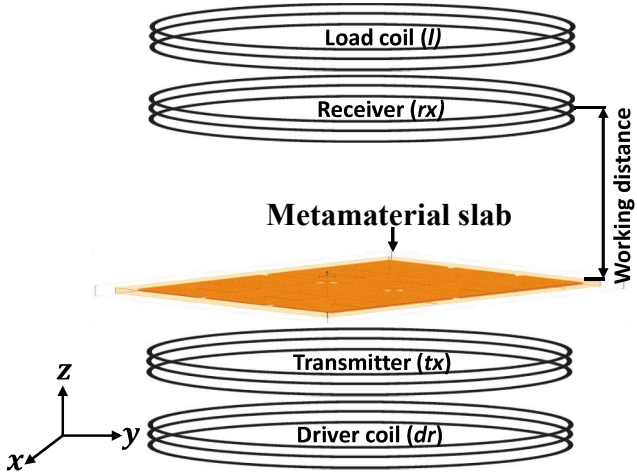

Fig. 3. ANSYS HFSS simulation of the proposed equivalent circuit model. The Model comprises a 3×3 periodic array of unit MM cells incorporated between the resonant coils. The dimension of the metamaterial slab is 128 × 96$mm^2$

The model is designed and simulated in ANSYS High-frequency Structure Simulator (HFSS) using the concept of FEA. Each of the drive coil, transmitter coil, receiver coil and load coil consist of three turns of copper coil while the metamaterial is fabricated with Ferrite material and litz wire. The choice of litz wire is predicated on its inherently low loss characteristics at high frequency, an important requirement for achieving high power transfer efficiency. The simulation parameters presented Table I are consistent with IEEE standard requirement which stipulate a criterion of 5% for ripple minimization [7], [21]. It is worth stating that for the entire simulation, the magnetic air-gap between the transmitting coil and MM-slab is varied from 100$mm$ to 250$mm$. The resulting magnetic field distribution based on simulation of the WPT system with and without the integration of MM-slab is depicted in Fig. 4(a) and Fig. 4(b), respectively. Apparently, the distribution and intensity of the magnetic field is more concentrated and about ×10 larger when MM-slab is inserted between the transmitter coil and receiver compared to conventional WPT system. The observed performance enhancement due to the inclusion of MM can be attributed to its evanescent wave amplification characteristics which essentially converges near magnetic field and leakage flux lines while amplifying and effectively focusing the traveling magnetic flux lines towards the receiver coil. It is worth stating that as the distance between the transmitter and receiver increases, the magnetic flux lines diverges leading to power dissipation and a diminution in overall power transmission efficiency.

TABLE I
PARAMETER SPECIFICATION OF EQUIVALENT CIRCUIT MODEL

| Parameter | Symbol | Unit | Value |
|---|---|---|---|
| Inductance | $L_{dr}, L_l$ | $mH$ | 0.7 |
| | $L_j$ | $\mu H$ | 1.49 |
| | $L_{tx}, L_{rx}$ | $\mu H$ | 4 |
| Capacitance | $C_{tx}, C_{rx}$ | $pF$ | 40 |
| | $C_i$ | $pF$ | 100 |
| Resistance | $R_s, R_l$ | Ω | 50 |
| | $R_{tx}, R_{rx}$ | Ω | 0.05 |
| | $R_i$ | $pF$ | 40 |

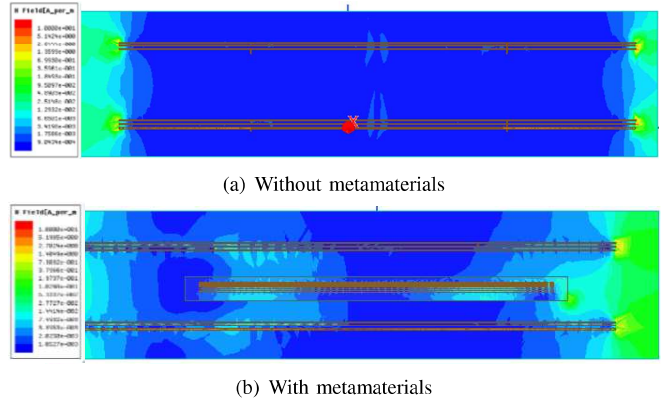

(a) Without metamaterials

(b) With metamaterials

Fig. 4. ANSYS HFSS WPT simulation showing the distribution of magnetic field in a four coil WPT structure (a) without metamaterial slab (b) with metamaterial slab

### IV. RESULTS AND DISCUSSION

#### A. Model Simulation Results and Analysis

Fig. 5 depicts the plot of power transfer efficiency against normalized transfer distance for a WPT system with and without the integration of MM. The normalized distance is based on the ratio of the maximum transfer distance/airgap between the receiver and MM-slab and individual transfer distance considered for the simulation. Conspicuously, the efficiency increases significantly for all instances of transfer distance when the WPT system is integrated with MM. The result validates the aforementioned concept of evanescent wave amplification and coupling of near magnetic field which is responsible for mitigating leakage magnetic flux and power

275

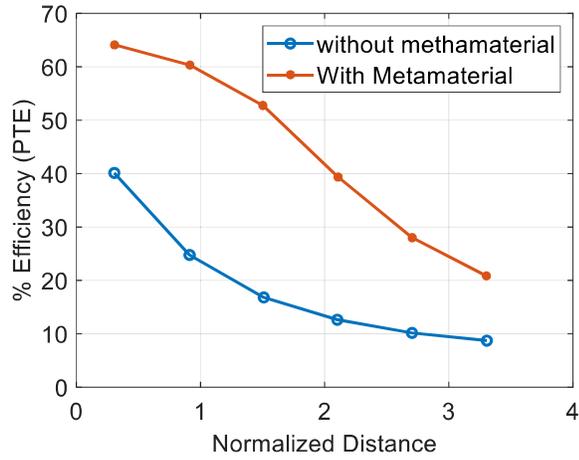

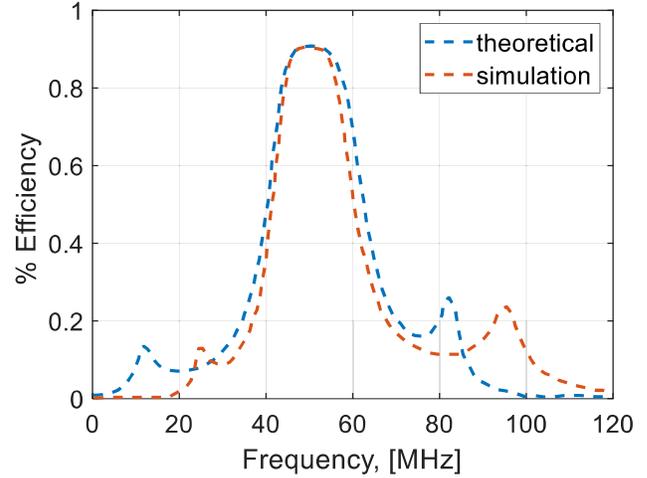

Fig. 5. Comparison of transfer efficiency of the WPT system with metamaterial and without metamaterial. The normalized distance is based on the ratio of parametric transfer distances and the maximum transfer distance considered for the full-wave simulation

Fig. 7. Waveform Indicating comparison of calculated and simulated power transfer efficiency. The simulation is based on FEA of the model in ANSYS

losses associated with conventional WPT system. In general, the forward transmission coefficient, $S_{21}$, represents the transfer efficiency. Using the concept of Finite Element Analysis (FEA) based on ANSYS HFSS, the simulated value of $S_{21}$ and PTE were extracted. Theoretically, the forward transmission scattering parameters, $S_{21}$ and the power transfer efficiency ($PTE$) can be evaluated based on (12) and (13), respectively. Going forward, $S_{21}$, which represents the efficiency of the system is obtained by varying the position of the MM-slab relative to the transmitting coils while taking measurement values at each position.

The simulation waveform depicting the magnitude of $S_{21}$ versus frequency for transfer distances ranging from $100mm$ to $250mm$ is shown in Fig. 6. When the transfer distance is fixed at $400mm$, and the distance between the MM and transmitter coil is varied from $100mm$ to $250mm$ , the resonance frequency is observed to vary between 12MHz-15MHz while the maximum value of $S_{21}$ corresponds to a resonant frequency, $f_o \approx$13.6 MHz. Obtrusively, the value of $S_{21}$ is closely related to the loading position of MM. Additionally, it is observed that when the MM is located away from the transmitting coil, the values of $S_{21}$ first increase and then decrease, corresponding to $S_{21}$ peaks ranging from 40% to 50%. It is also worth stating that the amplitude of $S_{21}$ attains its maximum when the MM is placed in the middle of the WPT system, indicating the possibility of achieving a better transmission efficiency by placing the MM in the middle of the four coil structure. Going forward, the comparison plot showing the simulated and calculated values of $S_{21}$ and PTE based on the equivalent circuit model is presented in Fig. 7. The waveform demonstrates a resonant operating frequency of 50MHz for both simulation and theoretical estimation, corresponding to 90% PTE. Similarly, the magnitude plot of $S_{21}$ against frequency for both simulation and theoretical analyzes is presented in Fig. 8. Although the simulation magnitude of $S_{21}$ is slightly larger than its calculated counterpart, the slight discrepancy in magnitude is suspected to be due to the inherent approximation in the analysis of the equivalent circuit. In addition, the variation of $S_{21}$ magnitude with transfer distance for both simulation and theoretical calculation is exhibited in Fig. 9. Based on the waveform, it is apparent that the simulation results closely match the theoretical calculation, effectively validating the feasibility of the proposed equivalent circuit model in verifying the transfer characteristics of the proposed MM-based WPT system. Further, it is notable that the magnitude of $S_{21}$ attains its maximum when the MM is situated at the middle of the WPT system. This represents the optimal position of the meta-material slab. Lastly, the comparison of power transfer efficiency of the four coil equivalent circuit is compared to those of two coil structure and capacitor-inductor-capacitor (CLC) resonant tank as demonstrated in Fig. 10. While a close correlation of the PTE is observed at the resonance peak for the three resonant tanks under consideration, it is apparent that the four coil structure has minimal resonance peak, essentially rendering it a preferable option for mitigating the electromagnetic leakage and total harmonic content while increasing the power transfer efficiency of the proposed MM-based WPT system.

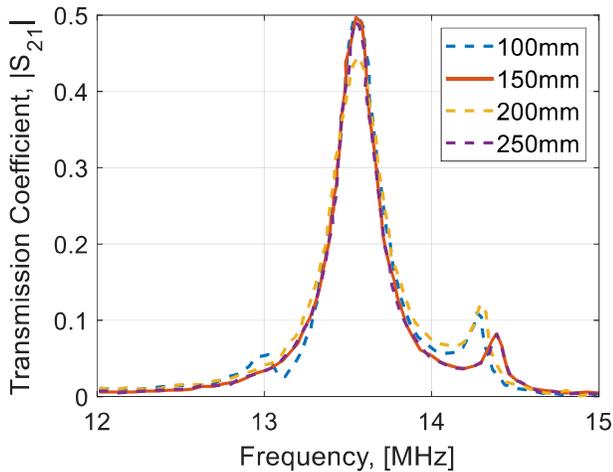

Fig. 6. Plot of forward transmission coefficient, $S_{21}$ as a function of Frequency for Varying values of transfer distance

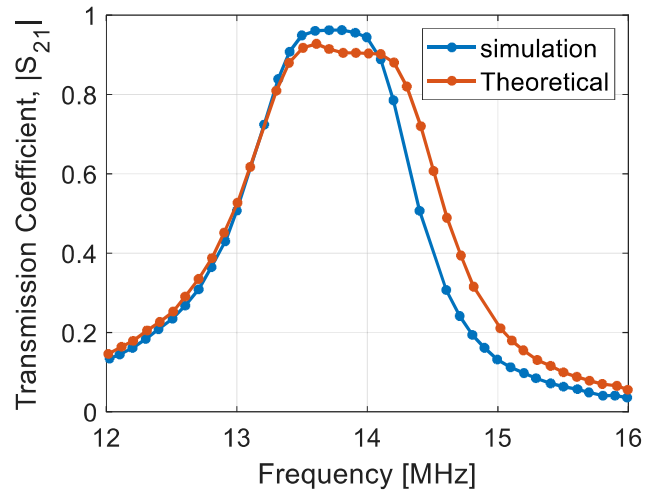

Fig. 8. Waveform Indicating Comparison of Calculated and Simulated Forward transmission coefficient ($S_{21}$) against frequency. The simulation is based on Finite Element Analysis of the Model in ANSYS HFSS

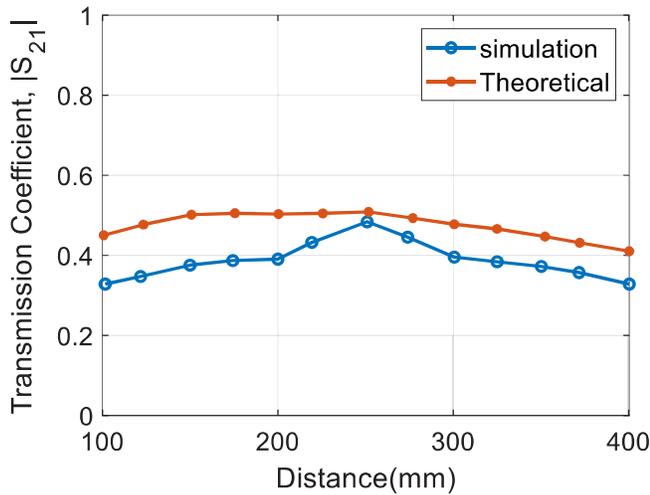

Fig. 9. Waveform Indicating Comparison of Calculated and Simulated Forward transmission coefficient ($S_{21}$) versus Transfer Distance. Model simulation is based on Finite Element Analysis in ANSYS HFSS

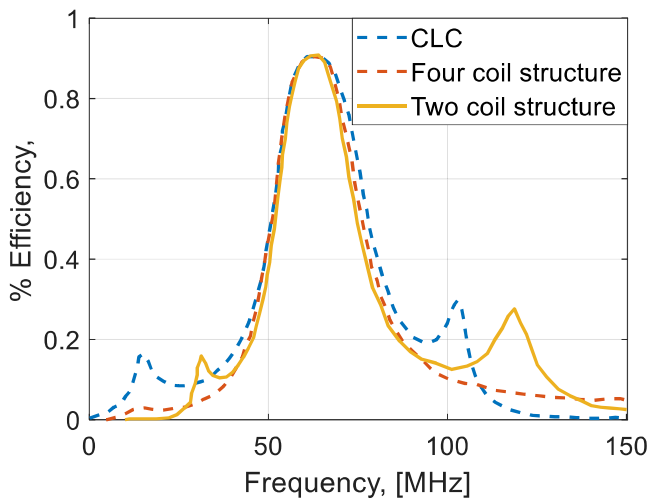

Fig. 10. Comparison waveform of power transfer efficiency based on capacitor-inductor-capaccitor (CLC), four coil structure and two coil structure.

## V. CONCLUSION

In this study, an equivalent circuit model is presented to simplify the analysis of an MM-based WPT system. In addition, a four-coil WPT system incorporating an MM-slab is set up. The circuit parameters are obtained by theoretical calculation and finite element simulation. The simulation results based on ANSYS HFSS shows that the proposed system can work effectively in the condition of magnetic resonance by integrating the MM slab which can make the power transfer efficiency to reach a maximum of 13.56MHz in different distance. The electrical parameters were obtained by numerical calculation and finite element simulation and then used to explore the transfer characteristic of the circuit matrix using MATLAB script. Finally, the comparison between simulation results and theoretical analysis has been carried out to verify the accuracy of the method, accordingly. To our best knowledge, this paper has profound significance for subsequent theoretical investigation of MM-based WPT system while offering the potential to drive future industrial application. Experimental test bed of the proposed structure is currently underway at SOLBART Laboratory of Tennessee Technological university and the obtained results will be decimated in future transaction of this publication.